\documentclass[12pt,a4paper]{article}
\usepackage{graphicx}
\usepackage{a4wide}
\usepackage{amsmath}

\begin{document}
\thispagestyle{empty}
 
\title{Gateways towards quasicrystals.}
\author{Peter Kramer,\\ Institut f\"ur Theoretische Physik der Universit\"at  T\"ubingen,\\ Germany.}
\maketitle

\section{Introduction.}
\label{sec:intro}

The experimental discovery of quasicrystals by D Shechtman, D Gratias, I Blech, and J W Cahn   in 1984  \cite{SH84}    provided the paradigm for a new type of long-range order 
of solid matter in nature. This discovery  stimulated an explosion  of new experimental and theoretical research. 
In years prior to the discovery, there was a very active development of various gateways to quasicrystals in theoretical and mathematical physics. Without this conceptual basis, it would have been
impossible to grasp and explore efficiently the structure and physical properties of quasicrystrals.
The aim in what follows  is to give a non-technical and condensed account of the conceptual gateways to quasicrystals
prior to their discovery.

\section{A Bravais, J B J  Fourier, A M Sch\"onflies and E S  Fedorov: Classical periodic crystallography.}
\label{sec:classical}
\begin{figure}[t]
\begin{center}
\includegraphics[width=0.4\textwidth]{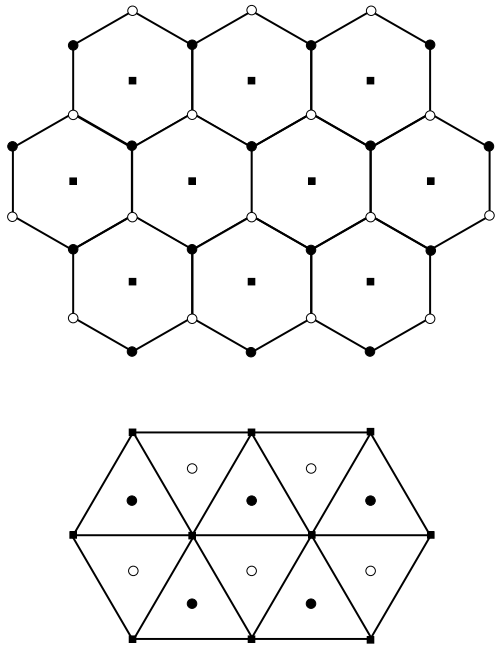}
\end{center}
\caption{\label{vorde} {\bf Hexagonal and dual triangular periodic lattices}.
The centers (white and black circles)  of the dual triangular cells  are located at the vertices of the hexagonal lattice cells (centers black squares). The hexagons and triangles are examples of the Voronoi and dual Delone cells in general periodic lattices.}
\end{figure}

Crystals in the natural world caught the attention by their regular geometrical
polyhedral form. It was A Bravais \cite{BR50}
who found the fundamental insight into their
internal structure by introducing the idea of an underlying periodic lattice $\Lambda$, see 
Fig. \ref{vorde}. 
Bravais was able to explain the regularity of crystal faces  by associating them with 
planes uniformly occupied by lattice points. 
The finite translations that connect lattice points form the translation group of the lattice,
denoted also for short by $\Lambda$.
If the next distances and directions in a lattice are tuned in particular ways, finite  rotations with respect to a fixed lattice point may turn lattice points into lattice points. The lattice then is compatible with a point group.
The combined symmetry under lattice translations and 
rotations led to the concept of space group symmetries. 
The question then arose: what are the possible crystal structures in 3-dimensional space?
This question was explored in the 19th century and culminated in the systematic classification 
of all possible crystal structures in terms of space group theory  by Sch\"onflies \cite{SC86/87} and Fedorov 
\cite{FE91}. The emergence  of an atomic structure of solid matter in the 19th century offered the possibility of viewing a crystal lattice as being formed by atoms. This structure was verified in 1912 in diffraction experiments with X rays following von Laue \cite{LA12} and Bragg \cite{BR12}.

The concept of a periodic lattice implies other basic notions for crystals: The periodic lattice symmetry 
requires that the long-range distribution of atoms is completely  determined once it is known  inside a unit cell. The analysis of periodic systems was fundamentally advanced by Fourier's \cite{FO22} concept
of the series expansion of a periodic function  into elementary periodic functions. 
For a complex-valued periodic function  $f^p(x)$ on the real line, this expansion in a condensed complex version takes the form
\begin{eqnarray}
 \label{qu1}
&&f^p(x)= \sum_{\nu} a(k_{\nu}) \exp (ik_{\nu} x),\: k_{\mu}= \frac{1}{2\pi} \mu, \mu= 0,\pm 1,\pm 2,...
\\ \nonumber
&& a(k_{\mu})= \int_{-1/2}^{1/2}dx\;  f^p(x) \exp (-ik_{\mu}x).
\end{eqnarray}
The Fourier coefficients $a(k_{\mu})$ in this expansion, given as integrals over the function $f^p(x)$  inside the unit interval, may be considered as functions defined on the points $k_{\nu}$ of a lattice $\Lambda^R$ in a Fourier $k$-space. Then the Fourier series represents the function $f^p(x)$ with domain the unit interval as a function $f^p(k)$ on the points of the so-called  reciprocal lattice $\Lambda^R$ in $k$-space. For crystals with lattices in 3-dimensional space $E^3$, the Fourier coefficients live on a $E^3$ $k$-space equipped with a  3-dimensional reciprocal lattice $\Lambda^R$.

With the advent of
scattering theory by quantum wave mechanics, von Laue \cite{LA12} and
Bragg \cite{BR12} related  the magnitude of the Fourier coefficients directly to the observed 
intensity of waves scattered from crystals. The intensity in scattering from crystals 
is characterized by sharp peaks in selected directions. In mathematical terms one speaks of 
a Fourier point spectrum.
The determination of the atomic structure of matter up to date 
is based on the interpretation of scattering data by Fourier series analysis.

The three related notions of a periodic lattice $\Lambda$, an atomic unit cell, and a Fourier series analysis characterize  crystals as periodic atomic long-range structures. 

\section{Point symmetry: Das Pentagramma macht Dir Pein?}
\label{sec:penta}

Another geometric 
aspect of crystals observed in nature were the systematic angles between 
their outer faces. With respect  to the center of the crystal, these faces often 
displayed 2fold, 4fold or 6fold point symmetry as part of their polyhedral symmetry.
These properties found an explanation in terms of the Bravais periodic lattice theory: 
It was shown   that all the observed  point symmetries could  be related  to   what became known as the seven Bravais lattices. 
The compatibility of point and periodic lattice symmetry in the framework of space groups 
formed the basis of the classification by Schoenflies and Fedorov.

There remained an enigma expressed by J W Goethe in his drama Faust \cite{GO08}:
{\em  Das Pentagramma macht Dir Pein?} 
Certain well-known point symmetries did not fit into 
lattice theory. Among them are the 5fold and the icosahedral symmetry, associated with the cyclic group $C_5$ and the icosahedral group ${\cal J}$ of rotations in 2- and 3-dimensional space. Already Plato \cite{PL91}
in his study of regular polyhedra had noted the regular dodecahedron and icosahedron
with icosahedral symmetry. For him, four   regular polyhedra were geometric building blocks  of the four elements whereas the dodecahedron he associated with the overall symmetry.

\begin{figure}[t]
\begin{center}
\includegraphics[width=0.5\textwidth]{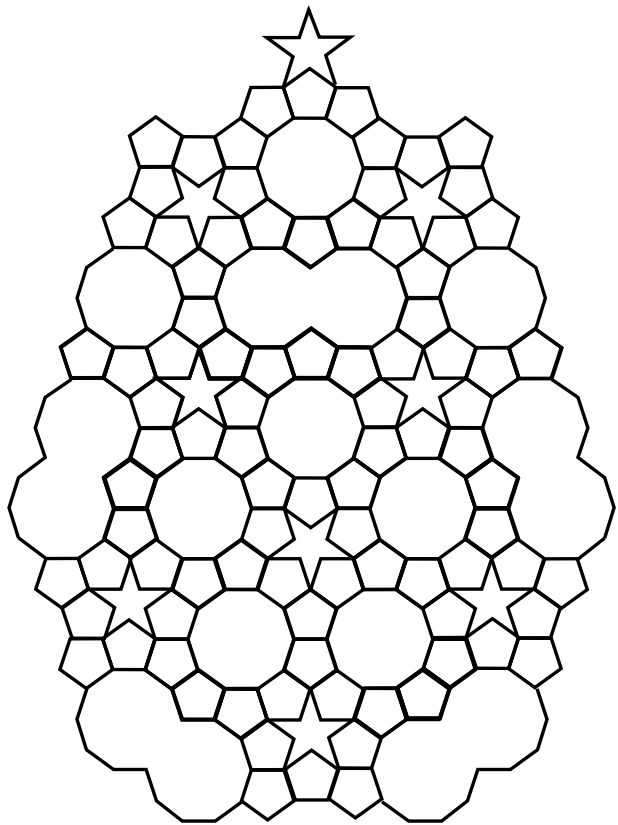}
\end{center}
\caption{\label{fig_kepler} {\bf Kepler's planar tiling with decagons and pentagons.} He found that, to leave no gaps, he needed pentagons and pentagonal stars in addition to decagons.}
\end{figure}

J Kepler \cite{KE38}  was impressed by the Platonic catalogue. In a first attempt he tried to use them
for the determination of the radii of spheres of the planets. Later, after his discovery of the elliptic 
orbits for the planets, he studied \cite{KE40} regular polygons, see Fig. \ref{fig_kepler}, and  polyhedra in order to deduce rational relations between 
astronomical data for the orbits of the planets. In his studies he also looked at tilings of the plane by regular pentagons, and enlarged the list of polyhedra by the half-regular triacontahedron.

With the success of crystallographic lattice theory  in the 19th century and thereafter 
its atomic setting by quantum theory, 5fold and icosahedral point symmetry, being incompatible with 
any 2- or 3-dimensional lattice, were stigmatized as being non-crystallographic.
Of course, the 5fold and icosahedral point symmetry can and does appear in molecules.
But all the known paradigms of long-range order,  thought to be periodic and so lattice-based,
excluded these point symmetries.

In view of their geometrical possibility  in 3-dimensional space the enigma remained: Are these point symmetries simply forbidden in nature, since they are not compatible with any lattice,
or can they be the gateways  towards a new type of long-range order in nature?

\section{H Zassenhaus and C Hermann 1948,49: Mathematical crystallography in $n$  dimensions for $n>3$.}
\label{sec:ndim}

The determination of all space groups in 3-dimensional  space was a clear classification problem in  mathematical physics. This classification obviously had a counterpart in Euclidean spaces of higher dimension.
The systematic analysis of these symmetries and lattices was advanced in particuar by H Zassenhaus \cite{ZA48} and by C Hermann \cite{HE49}.
This work, reviewed by Schwarzenberger \cite{SC}, showed that the counterparts of all essential findings of classical 
crystallography in $E^3$ can be found in $nD$ lattices. In $E^4$  the classification of space groups was completed in the work of Brown et al. \cite{BR78}. Of course the lattices in $E^4$  also include 5fold point symmetry.

The work on high dimensional crystallography gained new weight in physics with the advent of quasicrystals.

\section{R Penrose 1974: Aperiodic tilings of the plane.}
\label{sec:penrose}

\begin{figure}[t]
\begin{center}
\includegraphics[width=0.5\textwidth]{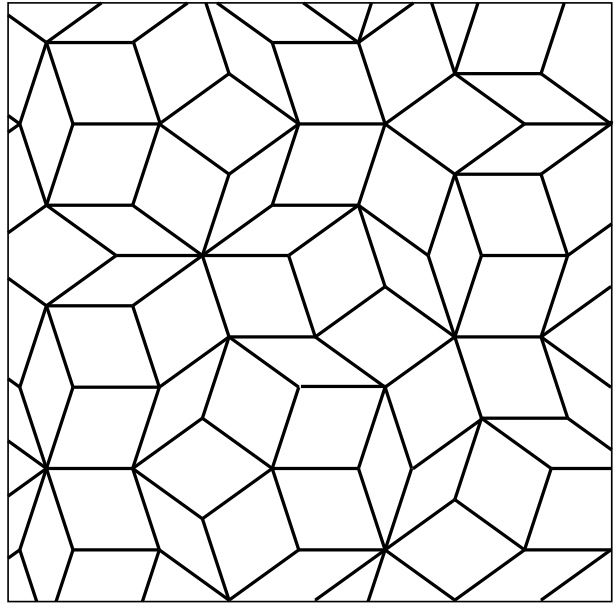}
\end{center}
\caption{\label{penq} {\bf The Penrose pattern}: The tiles have thin or thick rhombus shape.}
\end{figure}

In mathematical  crystallography, the Euclidean space $E^n$ is tiled without gaps or overlaps by repeated copies of the unit cell of the lattice. The position of the centers of these copies is given by all the lattice translations.
A natural generalization of periodicity are   tilings into copies of a finite number of cells.
If such a tiling cannnot be organized by a lattice, one has to find new ways to introduce a long-range order.
R Penrose \cite{PE74} proposed such a tiling of the plane with two rhombus tiles, known as the Penrose pattern, Fig. \ref{penq}. The edges of the two tiles have the same length. The angles between the edges of these 
tiles are multiples of $2\pi/5$ and so are adapted to 5fold symmetry. It follows that all the edges in the 
tile point in only five directions. This already suggests an average 5fold symmetry. 
Of course the tiles could be arranged into periodic tilings, but Penrose wanted to avoid a lattice periodicity. As a local rule for the long-range order he introduced the concept of matching rules. The matching rules demand that the marked 
directed edges of adjacent tiles must correspond to one another. 
Penrose demonstrated a number of interesting properties of his patterns.

The paradigm of the Penrose pattern was very appealing to scientists as a template for
a generalizations of classical crystallography. One could imagine to fix atoms to positions on the 
rhombus tiles and study the properties of the resulting generalized crystals.

Of particular interest was the question what other 
properties of classical crystallography extend to Penrose patterns.

A first question about the Penrose patterns was the  long-range order implied.
Can any patch of a tilings, built according to the local rules, be extended to cover the full plane? This 
question  has a negative answer: There are finite patches of tilings obeying the matching rules which cannot be extended in some parts without violating the matching rules. Penrose later in \cite{PE89}
called this the non-locality of the pattern.
Another problem were  the Fourier and diffraction properties of the Penrose pattern.

\section{A L Mackay 1981/2: Cells and diffraction properties from the Penrose pattern.}
\label{sec:mackay}

A L Mackay \cite{MA81}  presented the Penrose pattern as a paradigm for crystallography 
with 5fold point symmetry. He discussed the planar  Penrose rhombus pattern, suggested its two cells 
as non-periodic generalizations of crystallographic cells, and proposed the name quasi-lattice for the pattern. He  
also pointed out a 3D generalization to two rhombohedra whose edges point in the six directions 
perpendicular to the faces of the regular dodecahedron. He demonstrated in \cite{MA81} Figure 8  that these rhombohedra  can build Kepler's triacontahedron.

In \cite{MA82} he posed the question what diffraction would result if one placed 
scatterers to the vertices of the Penrose pattern. By an optical transform of circles, 
placed at vertex positions of a portion of a Penrose pattern, he arrived at a diffraction pattern 
governed by sharp peaks of intensity whose distribution  exhibited 10-fold point symmetry. 
Mackays result strongly suggested that the Penrose generalization of crystals shared with classical crystals the discrete point spectrum in diffraction which was the classical basis of structure determination by Fourier series analysis, section \ref{sec:classical}.

\section{H Bohr 1925: Quasiperiodicity and Fourier module from n-dimensional lattice embedding.}
\label{sec:bohr}

H Bohr back in the year 1925, in part II  of two papers \cite{BO25}, devoted to a a careful mathematical analysis  of almost periodicity, had  on  pp. 111-117, pp. 137-140, pp. 160-162  explored the notion 
of quasiperiodicity. He considered a lattice $\Lambda$ in a Euclidean space $E^n$ of dimension  $n>3$ and 
functions $f^p$ periodic on this space. His approach can be described as follows: He introduced a decomposition of Euclidean space $E^n= E^m_{\parallel}+E^{(n-m)}_{\perp}$, with  $E_{\parallel} \perp E_{\perp},$
into two orthogonal subspaces, chosen w.r.t. the lattice $\Lambda$ such that $E_{\parallel}$ was 
irrational. Irrationality meant: If $E^m_{\parallel}$ is parallel shifted by a vector $t$ so as to intersect with a lattice point $P$,
then the intersection of the shifted subspace with the set of all lattice point contains only $P$, 
$(E_{\parallel}+t) \cap \Lambda= P$.
The Fibonacci tiling, see section \ref{sec:fibon},  provides the simplest example of an irrational section.

Bohr then analyzed the restriction  of a periodic function $f^p$ on $E^n$, with domain restricted to the subspace $E_{\parallel}$.
He showed that this restriction has quasiperiodic properties. Moreover he considered the 
Fourier series  of $f^p$. He analyzed the Fourier transform of a function restricted to the irrational parallel subspace.
His finding can be expressed in terms of the  lattice $\Lambda^R$ reciprocal to the original lattice 
$\Lambda$ in $E^n$ in Fourier $k$-space: If he projects the points of the reciprocal lattice to the  parallel $k$-subspace
$E^m_{\parallel}$, the discrete set of these projections carries the Fourier coefficients of the quasiperiodic function. The discrete set of projected reciprocal lattice points forms what
in mathematical terminology is called a {\bf Z-module}. Its points by construction can be related 
in $E^m_{\parallel}$ by integer linear combinations of basis vectors of the reciprocal lattice, projected to this subspace.
In contrast to the reciprocal lattice points of periodic crystals, the points belonging the quasiperiodic module are discrete but become dense, that is come arbitrarily close, to any other point of the module.

The ingredients of Bohr's description  of quasiperiodic functions were then a periodic lattice $\Lambda$
in $E^n, n>3$, and an irrational subspace $E^m_{\parallel}$. On this basis Bohr provided a discrete 
Fourier module whose points carried the Fourier coefficients for quasiperiodic functions.

The Fibonacci paradigm, discussed  in the next section, provides  a simple 
example  of Bohr's  theory. For general applications of Bohr's ideas there remained a problem: The  points of a lattice form only a countable subset in $E^n$ leaving ample gaps for irrational subspaces. 
Among the infinite set of irrational subspaces,
what guideline can lead to a significant choice?

\section{Fibonacci 1202, M Lothaire 1983: Scaling and  the square lattice.}
\label{sec:fibon}

Leonardo de Pisa published in 1202 in Pisa the hand-written monograph {\em Liber abaci}. In it he presented 
his famous series of the integer Fibonacci numbers defined recursively by 
\begin{eqnarray}
\label{qu1b}
&& f_{n+1}=f_n+f_{n-1},
\\ \nonumber 
&&f_1=f_2=1, f_3=2, f_4=3, f_5=5,...
\end{eqnarray}
The Fibonacci numbers appear in mathematical combinatorics, M Lothaire \cite{LO83} p. 10,
as follows: Consider an alphabet $A=\{a,b\}$ and words formed recursively by the concatenation of letters
\begin{eqnarray}
 \label{qu1bb}
&&fi_{n+1}=fi_nfi_{n-1},\: n \geq 2,
\\ \nonumber
&&fi_1=b, fi_2=a, fi_3=ab, fi_4=aba, fi_5=abaab,... .
\end{eqnarray}
Counting the number of letters in successive words, called the word length $|fi_n|$, one finds
\begin{equation} 
\label{qu1bc}
|fi_1|=|fi_2|=1, |fi_{n+1}|=|fi_n|+|fi_{n-1}|=f_{n+1}.
\end{equation}
So the word length is a Fibonacci number. From the relative frequency of the letters $\{a,b\}$ in the 
words one can easily proof that the Fibonacci words cannot be periodic.
The Fibonacci words can be converted into an aperiodic tiling by interpreting the letters $(a,b)$ 
as intervals on the line of length $(1,\tau)$ respectively. 

\begin{figure}[t]
\begin{center}
\includegraphics[width=1.0\textwidth]{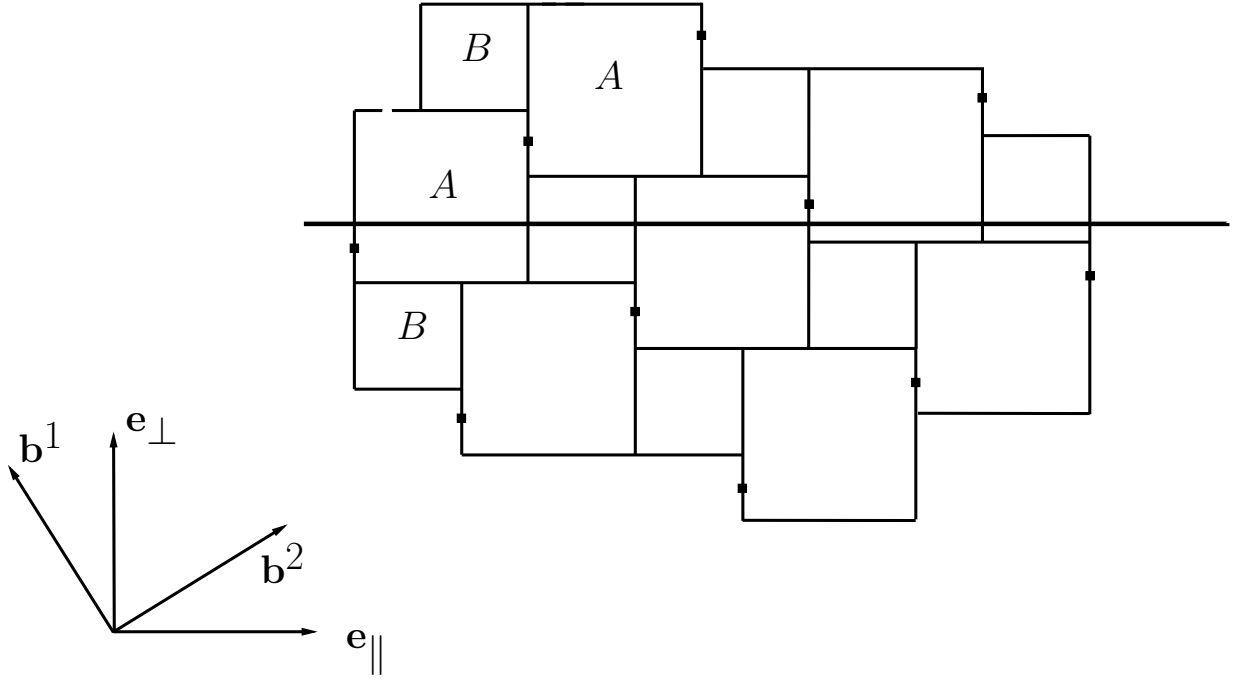}
\end{center}
\caption{\label{fiboq} The Fibonacci tiling from the square lattice. The black squares denote the 
points of the square lattice. The two squares $A,B$ give a periodic tiling of the plane. 
The Fibonacci matrix $g$ determines scalings in  two perpendicular directions 
$e_{\parallel}, e_{\perp}$.
A line parallel to $e_{\parallel}$ intersects the two squares in two intervals of length
in proportion $\tau$. The sequence of intervals  generates on the line the Fibonacci tiling,
beginning with $abaab$, eq. \ref{qu1bb}.}
\end{figure}

The recursive construction in eq. \ref{qu1bb} is the first  approach to the Fibonacci tiling.
In a second step we now relate the Fibonacci tiling to Bohr's theory of quasiperiodic functions. 
For this we follow \cite{KR06} pp. 311-12 and define 
\begin{equation}
 \label{qu1c}
g =
\left[ \begin{array}{ll}
1& 1\\
1&0\\
\end{array}
\right] 
\end{equation}
This matrix belongs to the group $GL(2,Z)$ with integer matrix elements and determinant $\pm 1$.
Computation of the powers of this matrix relates them to the Fibonacci numbers since 
\begin{equation}
\label{qu1d}
g^n=
 \left[ \begin{array}{ll}
f_n& f_{n-1}\\
f_{n-1}&f_{n-2}\\
\end{array}
\right],\: {\rm det}(g^n)=(-1)^n.
\end{equation}
We determine the two eigenvalues  of $g$ as 
\begin{equation}
 \label{qu1e}
\lambda_1= -\tau^{-1}=-\tau+1,\: \lambda_2=\tau= (1+\sqrt{5})/2.
\end{equation}
and get the eigenvectors from
\begin{equation}
\label{qu1f}
BgB^{-1}= 
\left[
\begin{array}{ll}
 -\tau^{-1}&0\\
0&\tau\\
\end{array}
\right],\: 
B= 
\left[
\begin{array}{ll}
 -\sqrt{\frac{-\tau+3}{5}}&\sqrt{\frac{\tau+2}{5}}\\
\sqrt{\frac{\tau+2}{5}}&\sqrt{\frac{-\tau+3}{5}}\\
\end{array}
\right]. 
\end{equation}
as the two orthonormal column vectors $B=:(b^1,b^2)$ of the matrix $B$.
These two vectors are obtained from an initial orthogonal basis by  application of the  matrix $B$.
Writing eq. \ref{qu1f}  as 
\begin{equation}
 \label{qu1g}
\left[
\begin{array}{ll}
 -\tau^{-1}&0\\
0&\tau\\
\end{array}
\right]B 
=B \left[ \begin{array}{ll}
1& 1\\
1&0\\
\end{array}
\right] 
\end{equation}
gives the following interpretation: 
The basis vectors $(b^1,b^2)$ span a square lattice. The integer linear combinations of the basis vectors $(b^1,b^2)$,  given  on the right-hand side by the right action 
of $g$ on $B$, transform lattice points into lattice points.
The left-hand side shows that the two vectors $(b^1,b^2)$ are 
scaled respectively  by the factors $(\lambda_1, \lambda_2)$. 
We can combine this result with the Bohr theory of quasiperiodic functions: The orthogonal basis vectors 
$e_{\parallel}=(1,0)\: e_{\perp}=(0,1)$  determine  two irrational orthogonal directions through the square lattice and provide one-dimensional irrational subspaces $E_{\parallel},E_{\perp}$. 
With respect to these vectors, the original matrix $g$ becomes diagonal.
One can construct \cite{KR06}
a new periodic tiling of $E^2$ by two squares whose boundaries run in the directions of these subspaces, see Fig \ref{fiboq}.

Now let a line parallel to $e_{\parallel}$ intersect these two squares in two intervals.
These intervals belong to  a module on the line. 
Their length 
is in the golden ration $\tau$. The tiling on the parallel line is the Fibonacci tiling.
A parallel scaling with factor $\tau$ transforms end points of intervals into end points.
So the Fibonacci tiling has a scaling symmetry and displays a cell structure. The scaling symmetry selects a particular one-dimensional
irrational subspace in the square lattice, and so by Bohr's theory becomes a source of quasiperiodicity. Similar lattice scalings by powers of $\tau$ appear in 
lattices of $E^4, E^6$ with 5fold and icosahedral symmetry. These scalings   underly the 
notions of inflation and self-similarity.

It follows from Bohr's theory that the Fibonacci tiling is quasiperiodic and 
has a discrete Fourier module. 

\section{ Y Meyer 1970, 1972: Harmonious sets.}
\label{sec:meyer}

In \cite{ME70}, \cite{ME72}, Y Meyer, starting from a mathematical study of harmonic analysis on locally compact abelian groups, 
introduced  certain discrete point sets he called harmonious sets. After the discovery of quasicrystals it was realized  that these harmonious or Meyer sets generalize the notion of lattices to in general aperiodic structures, and so provide a mathematical frame including and generalizing quasicrystals. 

Of  the seven equivalent characterizations of Meyer sets given by R V Moody \cite{MO97} pp. 403-41, we mention here only  a geometric one: 
It starts from a Delone set $\Lambda \in R^k$, defined as  a relatively dense and uniformly discrete set. 
This  Delone set becomes a Meyer set if there is  a finite set $F$ such that the  set of differences 
obeys $\Lambda-\Lambda= \Lambda+F$. Clearly one  sees the generalization from the notion of a lattice, 
whose set of differences would result in $F=0$.

R V Moody \cite{MO97} pp. 403-41 gives a detailed mathematical account   of Meyer's harmonious sets in the light of our present knowledge of aperiodic structures.
The broader field of mathematics for aperiodic structures 
is the subject of the  volume \cite{MO97}, edited in 1997 by R V Moody.

\section{ A Janner and T Janssen 1977: Fourier analysis of incommensurate and modulated crystals.}
\label{sec:incomodul}

A first and successful application in line with  Bohr's theory of quasiperiodic functions to crystallography was made by 
A Janner and T Janssen \cite{JA77}, \cite{JA78}, based on previous work of P M de Wolff \cite{WO72}. The idea was to describe so-called incommensurate 
and modulated structures, found in certain classes of crystals, by the extension of 3D space to a superspace equipped with a superlattice. The extra dimensions then are used  to describe 
incommensuration and modulation. 

Introduction of a reciprocal Fourier superspace, and projection to
the usual Fourier space then provided, beyond  the 
usual diffraction pattern, a pattern of sattellite  diffraction peaks whose structure encodes the 
specific nature of the incommensuration or modulation. 
Here the Fourier analysis beyond periodicity was developed, extended and applied in the spirit of Bohr's frame of quasiperiodicity.

\section{N de Bruijn 1981: The quasiperiodic Penrose pattern in 2 dimensions derived from a lattice.}
\label{sec:bruijn}

From the mathematical side, de Bruijn \cite{BR81} presented the first analysis of the Penrose pattern
by use of a lattice embedding into 4-dimensional space $E^4$. His choice of $E^4$ was guided by the wish  to incorporate 5fold point symmetry. The Euclidean space $E^4$ has the lowest dimension 
to allow for a lattice embedding with 5fold point symmetry. If one examines the action of the cyclic group $C_5$ on $E^4$, one finds two orthogonal subspaces of dimension $2$.
The unique subspace $E^2_{\parallel}$ of $E^4$ in which the cyclic group generator acts  as a rotation by an angle $2\pi/5$ is the natural choice of $E^2_{\parallel}$ for a quasiperiodic function. This subspace has the property of being 
irrational with respect to the chosen lattice in $E^4$.

The next task of de Bruijn was  to identify the Penrose 
rhombus tiles as projections. We take the liberty to describe  his finding  in the terminology of a later analysis of the same 
lattice in $E^4$ given in \cite{BA90}. The lattice provides  two tilings of $E^4$ by 4-polytopes: One is a tiling by Voronoi polytopes, which are 
the Wigner-Seitz cells of the lattice centered at the lattice points. The second, dual tiling is given by  so-called Delone polytopes, centered at the vertices of the Voronoi domains.
His geometric view allowed de Bruijn to identify the Penrose rhombus tiles with what is denoted in \cite{BA90} as the projections of 2-dimensional boundaries of the so-called Delone cells. De Bruijn  introduced a so-called pentagrid for the construction by projection of a Penrose rhombus tiling.

De Bruijn's  contribution to the theory of quasicrystal presented major advances: His construction 
of an irrational lattice embedding into 4 dimensional space 
related the planar Penrose tiling construction 
to the theory of Bohr. 
This construction showed that 
the requirement  of 5fold point symmetry uniquely determines  the irrational subspace 
required by Bohr's theory.  So indeed the 5fold point symmetry promised to be the gateway to a 
new type of long-range order.
When combined  with Bohr's theory, it followed from de Bruijn's construction that the Fourier transform 
of a Penrose pattern can be described by a module of sharp diffraction points, with positions 
the projections of 
the reciprocal lattice to a 2-dimensional Fourier k-subspace. This consequence  confirmed that Mackay's conjecture  of
sharp diffraction peaks from a Penrose pattern had a strict  mathematical basis.

\section{P Kramer 1982/84: Icosahedral tilings in 3 dimensions.}
\label{sec:kramer}

The pentagram  had led to the  Penrose paradigm of a quasiperiodic planar tiling in 2 dimensions.
Crystals in physics are phenomena in 3 dimensions.
The counterpart of the pentagram  is the icosahedron, whose point symmetry is forbidden in 3-dimensional lattices.  There arose now in 3 dimensional 
space
the question of tiles and tilings  with  forbidden icosahedral point symmetry. The first recursive and the second lattice approach discussed in section \ref{sec:fibon} for the Fibonacci tiling looked promising.
Kramer in \cite{KR82}  constructed a first set of seven 
elementary convex 
polyhedral tiles with two properties:\\
(i) Copies of them could be packed into a regular dodecahedron. \\
(ii)  Copies of the seven tiles could be packed into polyhedra of the same seven shapes, but scaled by 
a factor which was the third power of the golden section number $\tau=\frac{1}{2}(1+\sqrt{5})$. It was clear then that, by repeated 
application of this self-similar scaling, any region of 3-dimensional  space could be covered by a tiling of the 
seven elementary tiles. 
Mosseri and Sadoc \cite{MO82} managed to reduced the number  of these tiles 
from seven to four.

The findings by de Bruijn \cite{BR81} and by Bohr \cite{BO25} suggested the following lattice construction for icosahedral quasicrystals:\\ (1) One had to find a lattice in $E^n$ which under the action of the icosahedral group ${\cal J}$ is transformed into itself,
(2) Moreover one  should find  a subspace $E^3_{\parallel}\in E^n$ of dimension 3, invariant under the action of ${\cal J}$.\\
Kramer and Neri \cite{KR84} showed that the hypercubic lattice in 6-dimensional space was compatible with 
icosahedral point symmetry, and moreover provided a  unique 3-dimensional subspace  invariant under the icosahedral rotation group ${\cal J}$.

By considering the Voronoi 6-polytopes of the hypercubic lattice in $E^6$ and their 3-dimensional boundaries, both  projected to the parallel space $E^3_{\parallel}$ with icosahedral symmetry,
there emerged Kepler's triacontahedron from the Voronoi polytope, and rhombohedra in two shapes 
from the 3-dimensional boundaries. So this icosahedral tiling is organized exactly by the tiles considered by Mackay \cite{MA82}.  As was found out later, Kowalewski \cite{KO38} in 1938 in a book on recreational mathematics had already decribed 
Kepler's triacontahedron and the two rhombohedra as icosahedral projections of 
the hypercube in 6 dimensions  and its boundaries.

\begin{figure}[t]
\begin{center}
\includegraphics[width=0.4\textwidth]{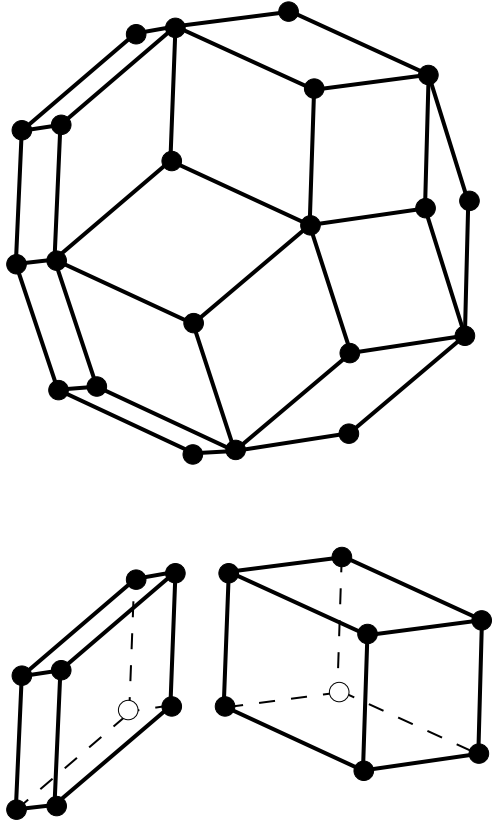}
\end{center}
\caption{\label{tiles1q} The icosahedral projection to $E^3_{\parallel}$ of the hypercubic Voronoi polytope in $E^6$ is Kepler's triacontahedron. The 3-dimensional boundaries of the hypercube project 
then into a thin and a thick rhombohedron. }
\end{figure}

This work generated  in three  dimensions the first  paradigm of a quasicrystal with  icosahedral point symmetry. Combined with Bohr's general 
theory, a diffraction analysis  on an appropriate icosahedral Fourier module could be devised.
A generalization  of de Bruijn's planar pentagrid to a hexagrid in 3-dimensional provided \cite{KR85}, \cite{KR85a} the construction of an icosahedral tiling from the two rhombohedral tiles. 
Duality in the high-dimensional lattice, in analogy of what is shown in Figure 
\ref{vorde}, plays a major role.

The enigma of the pentagram 
was finally solved, 5fold and icosahedral symmetry were back on their way into physics.

\section{D Shechtman, D Gratias, I Blech and J W Cahn  1984: Discovery of iscosahedral quasicrystals.}
\label{sec:shechtman}

In the previous sections we surveyed the theoretical approaches to quasicrystals prior to 
their experimental discovery. 

In 1984 D Shechtman, D Gratias, I Blech and J W Cahn \cite{SH84} announced the discovery of quasicrystals exhibiting a diffraction pattern with sharp peaks of icosahedral point symmetry. This discovery implied that atomic matter 
could organize itself in the new paradigm of quasiperiodic long-range order. 
An international  workshop at Les Houches in 1986 \cite{HO86} brought together many protagonists of quasicrystal
theory with D Shechtman and his collegues.
A brief review along similar lines as given here can be found in the epilogue by J W Cahn
\cite{JA95}, pp. 807-10 to the 5th International 
Conference on Quasicrystals, Avignon 1995.

\section{Postscriptum: D Levine and P J Steinhardt 1984, A Katz and M Duneau 1986, 
B Gr\"unbaum and G C Shepard 1987,
H Q Ye and K H Kuo et al. 1984, Ishimasa  et al.  1985.}
\label{sec:post}

The extraordinary development of quasicrystals after 1984,  both on 
the experimental and the theoretical level,    is a new story. 
Here it remains to  briefly postscribe theoretical and experimental  work by  authors
that was published  
shortly after the experimental discovery of quasicrystals. 

D Levine and P J Steinhardt \cite{LE84} in 1984 devised 
a construction method based on the Fibonacci sequence, and  
proposed the name quasicrystals for the new ordered structures. 
A Katz and M Duneau \cite{KA86} in 1986 developed  projection methods for the construction of icosahedral tilings 
by rhombohedra. Tilings were well described  in  a monograph written in 1987 by B Gr\"unbaum and G C Shepard
\cite{GR87}. 

Enlarging the field of quasicrystals on  the experimental side, H Q Ye and K H Kuo \cite{YE84}  in 1984 studied  quasicrystals with layers
of forbidden 10fold point symmetry. T Ishimasa, H U Nissen, and Y Fukano \cite{IS85} in 1985 prepared  structures 
with $(Ni, Cr)$ atomic composition and non-crystallographic 12fold point symmetry. 
 
After 1984, the broad development of quasicrystal preparation, structure analysis and new physical properties
became  manifest in the Proceedings of International Conferences on Quasicrystals, 1986 \cite{HO86} in Les Houches (France), 1989 \cite {YA90} in Vista Hermosa (Mexico),  1992 \cite{KEL93} in St Louis (USA), 1995 \cite{JA95} in Avignon (France), 1997 \cite{TA98} in Tokyo (Japan), and 1999 \cite{GA00} in Stuttgart (Germany).

\end{document}